\documentclass[letterpaper, 10 pt, conference]{ieeeconf}  
\IEEEoverridecommandlockouts
\usepackage{cite}
\usepackage{amsmath,amssymb,amsfonts}
\usepackage{algorithmic}
\usepackage{graphicx}
\usepackage{textcomp}
\usepackage{array}
\usepackage[mathscr]{euscript}
\usepackage{blindtext}
\usepackage{xcolor}
\definecolor{mblue}{rgb}{0,0.4470,0.7410}
\usepackage{mathtools}
\usepackage{arydshln}
\usepackage{subcaption}
\usepackage{xspace}

\newcommand{\R}{\mathbb{R}}

\newcommand{\M}{\mathscr{M}}
\newcommand{\N}{\mathbb{N}}

\newcommand{\dnu}{n_\mathrm{u}}
\newcommand{\dnx}{n_\mathrm{x}}
\newcommand{\dny}{n_\mathrm{y}}
\newcommand{\dnp}{n_\mathrm{p}}
\newcommand{\dnth}{n_\theta}
\DeclareMathOperator{\cayley}{Cayley}
\DeclareMathOperator{\tril}{tril}
\newcommand{\chris}[1]{{\color{mblue}#1}}

\newcommand{\ray}[1]{{\color{magenta}#1}}

\newcommand{\printnothing}[1]{}

\usepackage{amsthm}
\newtheorem{thm}{Theorem}
\newtheorem{prop}{Proposition}
\newtheorem{lemma}{Lemma}

\theoremstyle{remark}
\newtheorem{defn}{Definition}
\newtheorem{prob}{Problem}
\newtheorem{remark}{Remark}

\def\arXivversion{1}
\newcommand{\arXver}[2]{%
  \ifx\arXivversion\undefined%
    #1%
  \else%
    #2%
  \fi%
}

\begin{document}
\title{Learning Stable and Robust Linear Parameter-Varying State-Space Models}
\author{Chris~Verhoek, Ruigang~Wang and  Roland~T{\'o}th%
    \thanks{This work was partly supported by the European Research Council (ERC) under the European Union's Horizon 2020 research and innovation programme (grant agreement nr. 714663), the Eötvös Loránd Research Network (grant. number: SA-77/2021), and by the Australian Research Council together with the NSW Defence Innovation Network.}%
    \thanks{C. Verhoek and R. T\'oth are with the Control Systems Group, Eindhoven University of Technology, The Netherlands. R. Wang is with the Australian Centre for Robotics and the School of Aerospace, Mechanical and Mechatronic Engineering, The University of Sydney, Australia. R. T\'oth is also with the Vehicle Industry Research Center, Széchenyi István University, Hungary.}
    \thanks{C. Verhoek and R. Wang both contributed equally to this paper.} 
    \thanks{Corresponding author: C. Verhoek (\texttt{c.verhoek@tue.nl}).}%
}

\maketitle
\thispagestyle{empty} %

\begin{abstract}
This paper presents two direct parameterizations of stable and robust \emph{linear parameter-varying state-space} (LPV-SS) models. The model parametrizations guarantee a priori that for all parameter values during training, the allowed models are stable in the contraction sense or have their Lipschitz constant bounded by a user-defined value $\gamma$. Furthermore, since the parametrizations are \emph{direct}, the models can be trained using unconstrained optimization. The fact that the trained models are of the LPV-SS class makes them useful for, e.g., further convex analysis or controller design. The effectiveness of the approach is demonstrated on an LPV identification problem.
\end{abstract}

\section{Introduction}\label{s:introduction}
Systems in engineering are becoming more complex and are continuously being pushed to increase their efficiency, performance and throughput. This makes their behaviors becoming more and more dominated by nonlinearities, which makes the process of modeling these systems much more difficult, as modeling based on first-principles quickly becomes too tedious, costly, and/or inaccurate. Therefore, efficient data-driven modeling tools for these type of engineering systems are getting increasingly more important.

The class of \emph{linear parameter-varying} (LPV) systems has been established to provide a middle ground between the complex, but general, nonlinear system models and the easy-to-use, but rather limited, \emph{linear time-invariant} (LTI) system descriptions. In LPV systems, the signal relations are considered to be linear, just as in the LTI case. However, the parameters that define these relations are assumed to be functions of a measurable, time-varying signal -- the so-called \emph{scheduling variable} $p$, which captures the nonlinear/time-varying effects of the underlying system~\cite{Toth2010_book}. The linearity property of LPV systems makes them attractive for modeling, analysis and control and the framework is supported by extensions of many powerful approaches of the LTI framework.

LPV system identification methods \cite{Toth2010_book, CoxToth2021} have also matured to provide LPV surrogate models of systems based on data. However, despite the many advances, it has remained an open question whether it is possible to \emph{a priori} enforce stability and performance properties on the identified model. %
Despite the promising results that have been achieved for set membership identification based on LPV \emph{input-output} (IO) models \cite{cerone2012input} with a computationally intensive approach, the problem has remained  unsolved for other LPV model classes.

Over the years, \emph{deep-learning}-based system identification methods have been introduced for the data-driven modeling of complex nonlinear systems \cite{sjoberg1995nonlinear}, including methods that focus on LPV models \cite{verhoek2022deep, rizvi2018state, lachhab2008neural}. Generally, the \emph{recurrent neural network} (RNN) model structures, such as LPV-SS models with NN-based coefficient dependencies has been the main point of interest. This is because such models can provide efficient learning of the (often difficult to model) scheduling dependencies, significantly contributing to the accuracy and automation of the overall modeling process. However, the dynamic nature of RNNs implies that stability of the model plays a significant role in the training \cite{barabanov2002stability}. In modeling of general nonlinear systems with RNNs, this stability problem gained interest in recent years \cite{miller2018stable, fazlyab2019efficient, cohen2019certified} and lead to the developments of so-called implicit ANN network structures \cite{el2021implicit}, which allow for more systematic analysis. %
Based on this implicit structure, a major research effort has been spent on stability and performance analysis of dynamic neural network models~\cite{revay2020lipschitz, pauli2021training}, mainly based on Lipschitz and contraction \cite{Inc_Lohmiller1998} properties of the models. Although promising, many of these techniques require constrained optimization for the training of the networks, due to the enforced stability and/or performance constraint that increases the computational complexity. Inspired by this drawback, \emph{direct} parametrizations of robust and stable RNNs have been introduced in recent years \cite{wang2023direct, revay2023flexible}, which allow for learning stable and robust deep-learning-based nonlinear models using unconstrained optimization. 

In this work, we join the efficient and attractive properties of the LPV framework with the recently introduced direct parametrization approaches that can give a priori stability and performance guarantees. More specifically, as our main contributions, we propose two direct parametrizations of LPV-SS models with NN-based coefficients, which automatically guarantee that the LPV-SS model is stable in terms of contraction or have a prescribed bound on its Lipschitz constant. We achieve this by making use of the Cayley transform, which has been recently applied to achieve similar parametrizations for convolutional neural networks~\cite{pauli2023lipschitz}. The added value of the LPV-SS model structure is that the learned model could later be used for further analysis and controller design using the well-established tools of the LPV framework.  Moreover, we want to highlight that this a priori guaranteed stability and robustness property of the LPV-SS model is attractive to use in modeling problems where experiment-design is limited in terms of excitation range or impact on the production process (e.g., tank reactors in the process industry), while the model is expected to accurately describe the system behavior over the entire operating range.

To achieve this, first we introduce the problem setting %
 in Section~\ref{s:problem}, while the proposed solution, %
 i.e., our main result, is given in Section~\ref{s:mainresults}. 
We demonstrate the effectiveness of our results on an example in Section~\ref{s:example} and the conclusions are drawn in Section~\ref{s:conclusions}. 

\subsubsection*{Notation} 
$\mathbb{N}$ denotes the set of non-negative integers and $\mathbb{D}_+^n$  is the set of $n$-dimensional positive diagonal matrices. $\| \cdot \|_2$ denotes the Euclidian vector norm. For a matrix $A\in\R^{n\times n}$, %
$\tril(A)$ corresponds to %
the lower triangular part of $A$. Given a square matrix $M$ with $I+M$ invertible, its Cayley transform is defined as 
$\cayley(M):=(I-M)(I+M)^{-1}$. %

\section{Problem Statement}\label{s:problem}

{Given a data-set $\mathcal{D}_T:=\{u_t,p_t,\tilde{y}_t\}_{t=1}^T$ where $  u_t\in\R^{\dnu}$,  $p_t\in\R^{\dnp}$, $\tilde y_t\in\R^{\dny}$ are input, scheduling, and output signals of %
length $T\in\N$. We are interested in learning, i.e., identifying, a \emph{linear parameter-varying state-space} (LPV-SS) model $\M_\theta$ via 
\begin{equation}\label{eq:learning}
    \min_{\theta \in  \Theta } \quad \mathcal{L}(\M_\theta(u,  p), \tilde y)
\end{equation}
where $\mathcal{L}$ is the $\ell_2$-loss %
of the simulation error, i.e., $\sum_{t=1}^T \| \tilde{y}_t -y_t \|_2^2$, 
with $y=\M_\theta(u,  p)$ describing the forward simulated model response of $\M_\theta$ along the given input and scheduling trajectory $( u,  p)$ in $\mathcal{D}_T$ and estimated initial conditions.
The model $\M_\theta$ is described as %
\begin{equation}\label{eq:lpv}
	\begin{bmatrix}
	x_{t+1} \\ y_t
	\end{bmatrix}=
    \overset{W(p_t)}{\overbrace{
    \begin{bmatrix}
	A({p_t}) & B({p_t}) \\
	C({p_t}) & D({p_t}) 
	\end{bmatrix}
    }}
	\begin{bmatrix}
	x_t \\ u_t
	\end{bmatrix}+
	b({p_t}), %
\end{equation}
\printnothing{
\chris{Problem with the constant term $b_(p_t)$ and contraction, see Example~4 in \cite{Ruffer2013}.} 
\ray{This is not a big issue as $p_t$ usually belongs to a compact set and the mapping $b(p_t)$ is Lipschitz. Any feedforward DNN is Lipschitz-bounded w.r.t. some finite $\gamma$, even if one does not impose any Lipschitz constraint during the training. However, if we specify certain bound $\gamma$ on the mapping $p\rightarrow b$, then we may have better generalization results.}
}%
where $x_{t}\in\R^{\dnx},u_{t}\in \R^{\dnu},y_{t}\in\R^{\dny},p_{t}\in\mathbb{P}\subseteq \R^{\dnp}$ are the state, input, output and scheduling signals at time-instant $t\in\N$, respectively. {Here the actual functional dependency on the scheduling $p_t$ of the matrices $A(p_t), \dots, D(p_t)$ and of $b(p_t)$, which is a possible bias (trimming term), are collected into the function $\psi_\theta$ (see Fig.~\ref{fig:sched-map}). The function %
\begin{equation}
    \psi_\theta: p\in\mathbb{P} \mapsto \{W,b\}, %
\end{equation}
is considered as a \emph{deep neural network} (DNN) parametrized with} $\theta\in\R^{\dnth}$, which correspond to the learnable parameters.
This construction of the LPV model allows for a flexible choice of the dependency structure in $A(p_t), \dots, D(p_t), b(p_t)$, for instance, one can learn {an affine} scheduling relationship
\begin{equation}\label{eq:linear-schedule}
    \begin{bmatrix}
    \mathrm{Vec}(W(p_t)) \\ b(p_t)
\end{bmatrix}=\psi_\theta(p_t):= S_1p_t + S_0,
\end{equation}
with $\theta=(S_0, S_1)$ as the learnable parameters. In~\cite{verhoek2022deep}, $\psi_\theta$ is considered as a linear mapping while a $\mu(u_t,u_{t-1},\ldots,y_t,y_{t-1},\ldots)$ is learnt with a deep-neural network to synthesize the scheduling signal from input-output signals directly as $p_t=\mu(u_t,u_{t-1},\ldots,y_t,y_{t-1},\ldots)$. In this paper, we consider the scheduling signal to be given and being part of the data-set $\mathcal{D}_T$.

Furthermore, for the sake of simplicity, we consider \eqref{eq:lpv} without a dedicated noise model, under the assumption that the data-generating system has an \emph{output-error} (OE) type of noise structure. Note that estimation under an innovation noise model can be easily incorporated into \eqref{eq:learning}, see~\cite{verhoek2022deep}, and the results of the paper can be easily generalized to the resulting model structure.
\begin{figure}[!bt]
    \centering
    \includegraphics[scale=1]{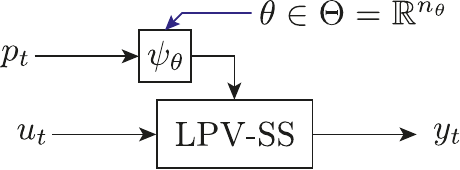}
    \caption{The LPV state-space model and its parameterized scheduling dependency $\psi_\theta$.} \vspace{-6mm}
    \label{fig:sched-map}
\end{figure}

In many applications, it is highly desirable to learn LPV-SS models via \eqref{eq:learning} with stability and robustness guarantees. Especially with a DNN parametrization of the coefficient functions, models estimated along the trajectory $\mathcal{D}_T$ tend to provide deteriorated performance and even unstable behavior when the scheduling trajectory leaves the region where $\mathcal{D}_T$ was obtained, causing much concern in their utilization for industrial applications. To prevent such phenomena occurring, we aim to ensure the following strong notions, which help the model to exponentially forget the initial conditions and generalize to unseen data {in a robust and stable manner:}

\begin{defn}\label{def:contr}
The system represented by %
\eqref{eq:lpv} is said to be \emph{contracting} if for any two initial conditions $ x_0^a,x_0^b\in\R^{\dnx} $, any bounded sequences $p\in\mathbb{P}^\mathbb{N}$, $u\in(\mathbb{R}^{\dnu})^\mathbb{N}$, the corresponding state sequences $ x^a, x^b $ satisfy
\begin{equation}
    \|x_t^a-x_t^b\|_2\leq K\alpha^t\|x_t^a-x_t^b\|_2,\quad \forall t\in \N,
\end{equation}
for some $K>0$ and $\alpha\in (0,1)$.
\end{defn}

\begin{defn}
    The system represented by \eqref{eq:lpv} is said to be \emph{$\gamma$-Lipschitz} for some $\gamma>0$, if for any initial state $ x_0\in\R^{\dnx} $, bounded parameter sequence $p\in\mathbb{P}^\mathbb{N}$, and bounded input sequence pair $(u^a,u^b)\in(\mathbb{R}^{2\dnu})^\mathbb{N}$, the corresponding output pair $(y^a,y^b)$ satisfies
    \begin{equation} \label{eq:gamma:L}
	\sum_{t=0}^T\|y_t^a-y_t^b\|_2^2\leq \gamma^2 \sum_{t=0}^{T} \|u_t^a-u_t^b\|_2^2,\quad \forall T\in \N.
    \end{equation} 
\end{defn}
Using these definitions, we solve the following problems in this paper:
\begin{prob}
Construct the model parameterizations 
\begin{subequations}\label{eq:modelparametrizations}
\begin{align}
        \M^c&:=\{ \M_\theta\mid \M_\theta \text{ is contracting } \forall \theta \in \R^{\dnth}\}, \\
        \M^\gamma&:=\{ \M_\theta\mid \M_\theta \text{ is $\gamma$-Lipschitz } \forall \theta \in \R^{\dnth}\}.
    \end{align} \end{subequations}
\end{prob}
\begin{remark}
    The following remarks are important:
    \begin{itemize}
        \item With the parameterizations in~\eqref{eq:modelparametrizations}, the learning problem~\eqref{eq:learning} can be formulated as an unconstrained optimization problem that can be solved by off-shelf first-order methods (e.g., stochastic gradient descent). This is because $\theta\in\Theta = \R^{\dnth}$.
        \item The $\gamma$-Lipschitz property is equivalent to having an incremental $\ell_2$-gain bound of $\gamma$ on \eqref{eq:lpv}, which in turn \emph{implies} an $\ell_2$-gain bound of $\gamma$ on \eqref{eq:lpv}~\cite{verhoek2023convex}.
        \item See also \cite{Inc_Lohmiller1998} for the connections between contraction and incremental stability.
    \end{itemize}
\end{remark}

\section{Main Results}\label{s:mainresults}
In this section, we first give sufficient conditions for contracting/$\gamma$-Lipschitz LPV-SS models and then present a direct parameterization such that those conditions are automatically satisfied {during training.}
\subsection{Stable and robust LPV-SS models}
To study the contracting or $\gamma$-Lipschitz property of \eqref{eq:lpv}, we first consider the error dynamics between two arbitrary trajectories of \eqref{eq:lpv} with the same scheduling signal, i.e., $(u^a, x^a,y^a,p)$ and $(u^b, x^b,y^b,p)$. For these trajectories, the error dynamics are: %
\begin{equation}\label{eq:err-dyn}
    \begin{bmatrix}
        \Delta x_{t+1} \\ \Delta y_t
    \end{bmatrix}=
    \begin{bmatrix}
	A({p_t}) & B({p_t}) \\
	C({p_t}) & D({p_t}) 
	\end{bmatrix} 
    \begin{bmatrix}
        \Delta x_t \\ \Delta u_t
    \end{bmatrix},
\end{equation}
where $\Delta x=x^a-x^b$, $\Delta u=u^a-u^b$ and $\Delta y=y^a-y^b$. %
Then, \eqref{eq:lpv} is contracting if \eqref{eq:err-dyn} is exponentially stable, while \eqref{eq:lpv} is  $\gamma$-Lipschitz  if \eqref{eq:err-dyn} has an $\ell_2$-gain bound of $\gamma$.
\begin{prop}\label{prop:conditions}
    The LPV-SS model \eqref{eq:lpv} describes a contracting system, if there exist a $\mathcal{X}\succ 0$ and an $\alpha\in (0,1]$ s.t. 
    \begin{equation}\label{eq:contraction}
        \alpha^2 \mathcal{X}-A^{\!\top}\!(p) \mathcal{X} A(p) \succ 0,\quad \forall p\in \mathbb{P}.    \end{equation}
    The %
    system is $\gamma$-Lipschitz, if there exist a $\mathcal{X}\succ 0$ s.t.
    \begin{equation}\label{eq:lipschitz}
        \begin{bmatrix}
            \mathcal{X} & 0 \\
            0 & \gamma^2 I
        \end{bmatrix}-
        W^{\!\top}\!(p)
        \begin{bmatrix}
            \mathcal{X} & 0 \\
            0 & I
        \end{bmatrix}
        W(p)
        \succ 0,\quad \forall p\in \mathbb{P}. %
    \end{equation}
\end{prop}
\begin{proof} %
Contraction of the system represented by \eqref{eq:lpv} is defined for the differential state under the same input sequence, hence \eqref{eq:err-dyn} with $\Delta u_t=0$ becomes
\begin{equation}
\Delta x_{t+1}=A(p_t)\Delta x_t.
\end{equation}
Based on \eqref{eq:contraction}, we have
\begin{equation}
    \alpha^2 V(\Delta x_t)\geq V(\Delta x_{t+1}),
\end{equation}
where $V(\Delta x)=\Delta x^\top \mathcal{X} \Delta x$, showing exponential (Lyapunov) stability of the error dynamics. This implies that the corresponding LPV-SS model is contracting.

To prove the $\gamma$-Lipschitz property of \eqref{eq:lpv}, we first multiply \eqref{eq:lipschitz}  from the left and right with $ \begin{bmatrix}
    \Delta x_t^\top  & \Delta u_t^\top 
\end{bmatrix}$ and $\begin{bmatrix}
    \Delta x_t^\top & \Delta u_t^\top
\end{bmatrix}^\top$, respectively. This leads to 
\begin{equation}
    \gamma^2 \|\Delta u_t\|_2^2 -\|\Delta y_t\|_2^2\geq V(\Delta x_{t+1})-V(\Delta x_t).
\end{equation}
Using a  telescoping sum based on the above inequality and that $\Delta x_0=0$, \eqref{eq:gamma:L} is satisfied. %
\end{proof}

\subsection{Model parameterization via Cayley transform}

The challenge in estimating $\psi_\theta$ and ensuring stability of~\eqref{eq:lpv} is that Condition~\eqref{eq:contraction} needs to hold for all $p\in \mathbb{P}\subset \R^{\dnp}$, representing an infinite-dimensional constraint that is required to be added to \eqref{eq:learning}. While it is possible to achieve some relaxation of this constraint, e.g., by restricting $\psi_\theta$ to be linear and $\mathbb{P}$ to a convex polytope and turn \eqref{eq:contraction} to a finite \emph{semi-definite programming} (SDP) problem, such relaxations (i) seriously restrict the representable class of systems and (ii) still involve a  significant amount of computation time, which can quickly make the training intractable. 
We tackle those issues by deriving an analytic solution to \eqref{eq:contraction}. 

\begin{thm}\label{thm:contracting}
The model \eqref{eq:lpv} defined by coefficient function $\psi_\theta$ satisfies \eqref{eq:contraction}, if and only if there exist $d\in \R^{\dnx}$, $\alpha\in (0,1]$, $\mathcal{Y}\in \R^{\dnx\times \dnx} $ and a mapping $\phi:p\mapsto (X,Y)$ with $X(p),Y(p)\in \R^{\dnx\times \dnx}$ such that
\begin{equation}\label{eq:Ap}
    A(p)=\alpha Q\Lambda^{-1}M(p) \Lambda Q^\top,
\end{equation}
with $\Lambda=\mathrm{diag}(e^{d})$ and 
\begin{equation}
    \begin{split}
        Q=\cayley(\mathcal{Y}-\mathcal{Y}^\top),\quad
        M(p)=\cayley(N(p)),
    \end{split}
\end{equation}
where $N(p)=X^{\!\top}\!(p) X(p)+Y(p) -Y^{\!\top}\!(p)+\epsilon I$ for some small positive constant $\epsilon$.
\end{thm}
\begin{proof} We first show that \eqref{eq:contraction} $\Leftrightarrow$ \eqref{eq:Ap} and then we prove that the invertible mapping between $\Lambda, Q, M(p)$ and $d, \mathcal{Y}, X(p), Y(p)$ can be easily established based on Lemmas~\ref{lem:cayley} and \ref{lem:cayley-2}, which are given in the Appendix. For the sake of notational simplicity, we use subscript $p$ to denote the dependency on the scheduling variable.

We first show that \eqref{eq:Ap} $\Rightarrow$ \eqref{eq:contraction}. By taking $\mathcal{X}=Q\Lambda^2Q^\top$, $\mathcal{X}\succ 0$ as $QQ^\top=I$ due to Lemma~\ref{lem:cayley-2}. Then, 
\begin{equation}
    \alpha^2 \mathcal{X}-A_p^\top \mathcal{X} A_p=\alpha^2 Q\Lambda(I-M_p^\top M_p)\Lambda Q^\top \succ 0,
\end{equation}
where positive definiteness of $I-M_p^\top M_p$ follows by Lemma~\ref{lem:cayley}.
Next, we show \eqref{eq:contraction} $\Rightarrow$ \eqref{eq:Ap}.
Since $\mathcal{X}\succ 0$, its \emph{singular value decomposition} (SVD) has the form $\mathcal{X}=Q\Sigma Q^\top $ with $\Sigma\in\mathbb{D}_+^{\dnx}$ and $Q^\top Q=I$, and $Q$ cannot have $-1$ as an eigenvalue. By letting $\Lambda =\Sigma^{1/2}$, we have that
\begin{equation}
    \begin{split}
        \eqref{eq:contraction} \Rightarrow I - M_p ^\top M_p \succ 0,
    \end{split}
\end{equation}
where $M_p=\tfrac{1}{\alpha} \Lambda Q^\top A_p Q \Lambda^{-1}$, which gives \eqref{eq:Ap}. \end{proof}

Thm.~\ref{thm:contracting} reveals that we can represent any $\psi_\theta$ coefficient function parametrization for which the defined model \eqref{eq:lpv} satisfies \eqref{eq:contraction} by the parameters $d, \mathcal{Y}$ and unconstrained mapping 
\[
\phi_{\tilde{\theta}}:p\mapsto (X, Y, B,C,D,b),
\]
which can be chosen as a DNN parametrized in $\tilde{\theta}$. This means that we can transform the learnable parameters $\theta$ to new parameters $\{d, \mathcal{Y}, \tilde{\theta}\}$ that guarantee that, for any value of them, the corresponding model \eqref{eq:lpv} satisfies \eqref{eq:contraction}.

In fact, we can use any parameterization for $\phi_{\tilde{\theta}}$, like a simple linear mapping \eqref{eq:linear-schedule}, or a polynomial parametrization, etc. This underlines the applicability of Thm.~\ref{thm:contracting}  beyond deep-learning-based identification of LPV models. %
Similar results can be derived for the $\gamma$-Lipschitz property.
\begin{thm}\label{thm:lipschitz} The model \eqref{eq:lpv} defined by coefficient function $\psi_\theta$ satisfies \eqref{eq:lipschitz}, if and only if there exist $d\in \R^{\dnx}$, $\mathcal{Y}\in \R^{\dnx\times \dnx}$ and a mapping $\phi:p\mapsto (X,Y,Z)$ with $X(p),Y(p)\in \R^{n\times n}$ and $Z(p)\in \R^{n_0\times n}$, where $n=\dnx+\min(\dnu,\dny)$ and $n_0=|\dny-\dnu|$, such that 
\begin{equation}\label{eq:Wp}
    W(p)=\begin{bmatrix}
    Q \Lambda^{-1} & 0 \\ 0 &  I
\end{bmatrix} M(p) \begin{bmatrix}
    \Lambda Q^\top & 0 \\ 0 & \gamma I
\end{bmatrix},
\end{equation}
with 
\begin{equation}\label{eq:extend-cayley}
    \begin{bmatrix}
        \cayley(N(p)) \\
        -2Z(p)(I+N(p))^{-1}
    \end{bmatrix}=\begin{cases}
    M(p), &\!\!\! \text{if } \dny\geq \dnu, \\
    M^{\!\top}\!(p), &\!\!\! \text{if } \dny< \dnu,
    \end{cases}
\end{equation}
where $N(p)=X(p)^\top X(p)+Y(p) - Y(p)^\top+Z(p)^\top Z(p)+\epsilon I$ with $\epsilon$ as a small positive constant.
\end{thm}
\begin{proof}
We first rewrite \eqref{eq:lipschitz} as follows
\begin{equation}\label{eq:lmi}
    \mathcal{X}_\gamma- W_p^\top \mathcal{X}_I W_p\succ 0
\end{equation}
where $\mathcal{X}_\gamma=\mathrm{diag}(\mathcal{X},\gamma^2 I)$ and $\mathcal{X}_I=\mathrm{diag}(\mathcal{X},I)$. By taking the SVD decomposition $\mathcal{X}=Q\Sigma Q^\top\!$ and letting $\Lambda=\Sigma^{1/2}$, we have $I-M_p^\top M_p\succ 0$ where 
\begin{equation}
    M_p = \begin{bmatrix}
        \Lambda Q^\top & 0 \\ 0 & I
    \end{bmatrix} W_p \begin{bmatrix}
        Q\Lambda^{-1} & 0 \\ 0 & \gamma^{-1}I
    \end{bmatrix}.
\end{equation}
Then, the techniques used in the proof of Thm.~\ref{thm:contracting} can be directly applied to prove \eqref{eq:lipschitz} $\Leftrightarrow$ \eqref{eq:Wp}.
\end{proof}
\begin{remark}
    The transformation in \eqref{eq:extend-cayley} can be considered as the Cayley transform for non-square matrices. When $\dny=\dnu$, the normal Cayley transform is recovered, as in that case, $Z(p)$ is an empty matrix.
\end{remark}

\section{Example}\label{s:example}
\newcommand{\testa}{\textsc{Test-a}\xspace}
\newcommand{\testb}{\textsc{Test-b}\xspace}
\newcommand{\train}{\textsc{Training}\xspace}
\newcommand{\valid}{\textsc{Validation}\xspace}
\begin{figure*}
\newlength\removewhitespaceforfig\setlength\removewhitespaceforfig{-4mm}
    \centering
    \begin{minipage}[t]{0.245\linewidth}
       \includegraphics[width=\linewidth]%
       {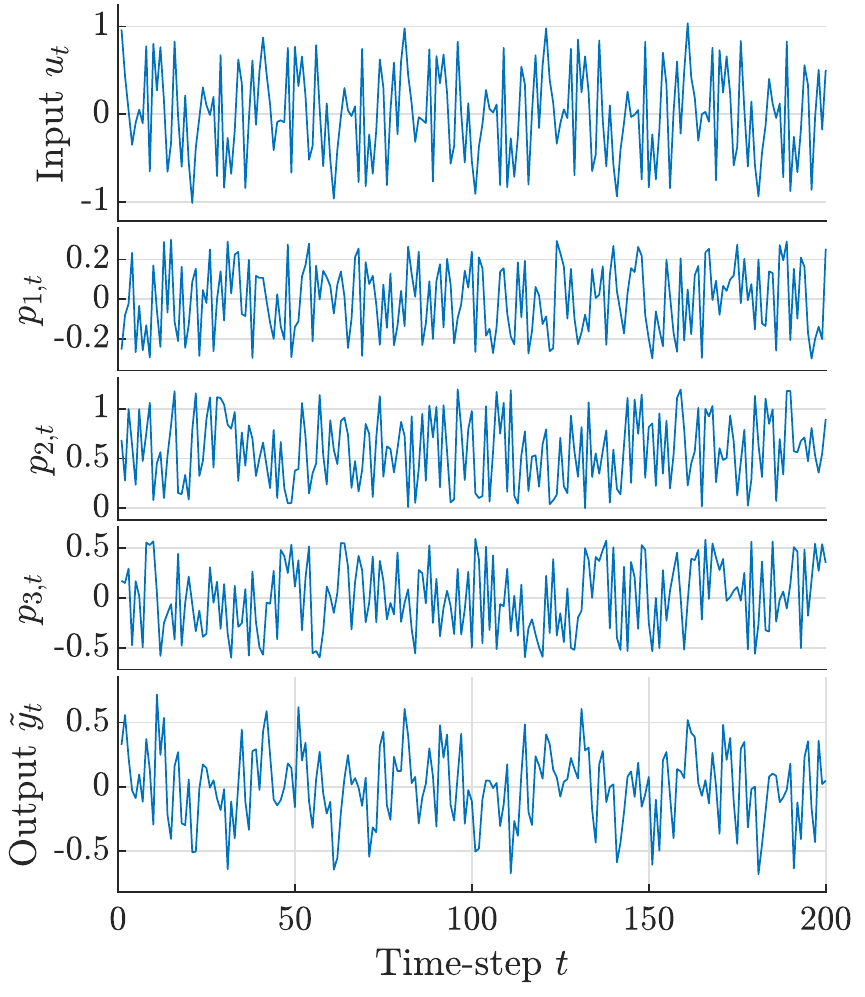}
       \caption{\train data-set}
       \label{fig:data1}
    \end{minipage}\hfill
    \begin{minipage}[t]{0.245\linewidth}
       \includegraphics[width=\linewidth]%
       {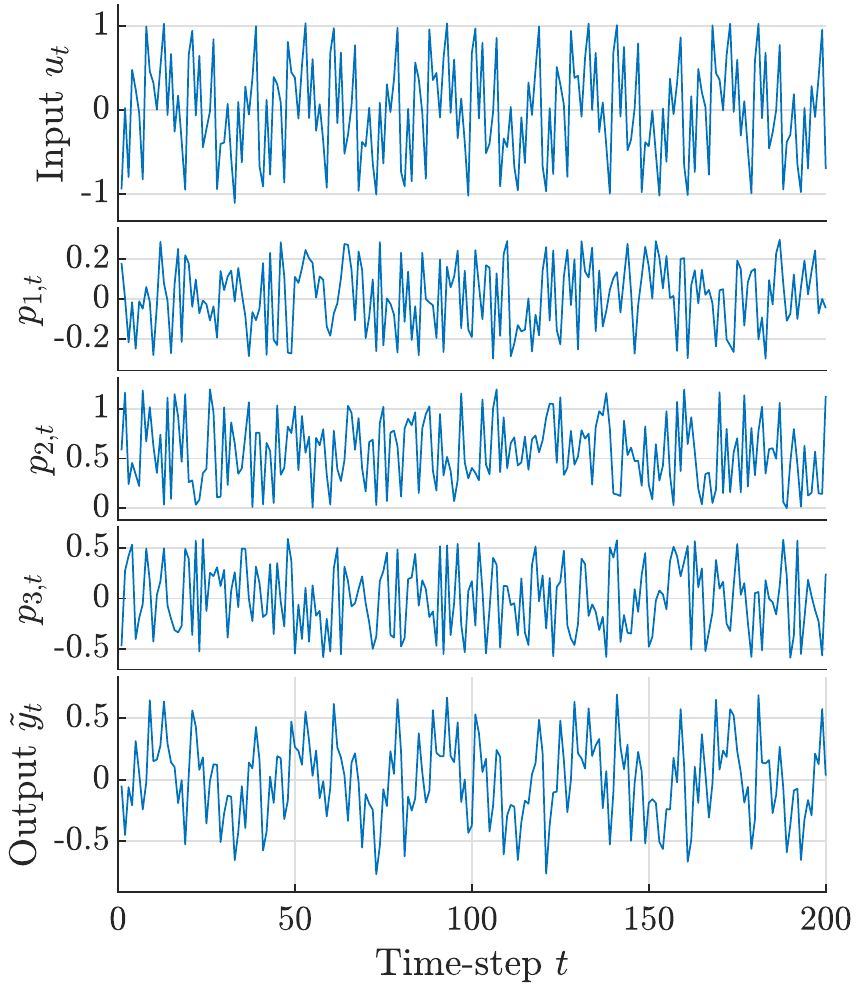}
       \caption{\valid data-set}
       \label{fig:data2}
    \end{minipage}\hfill
    \begin{minipage}[t]{0.245\linewidth}
       \includegraphics[width=\linewidth]%
       {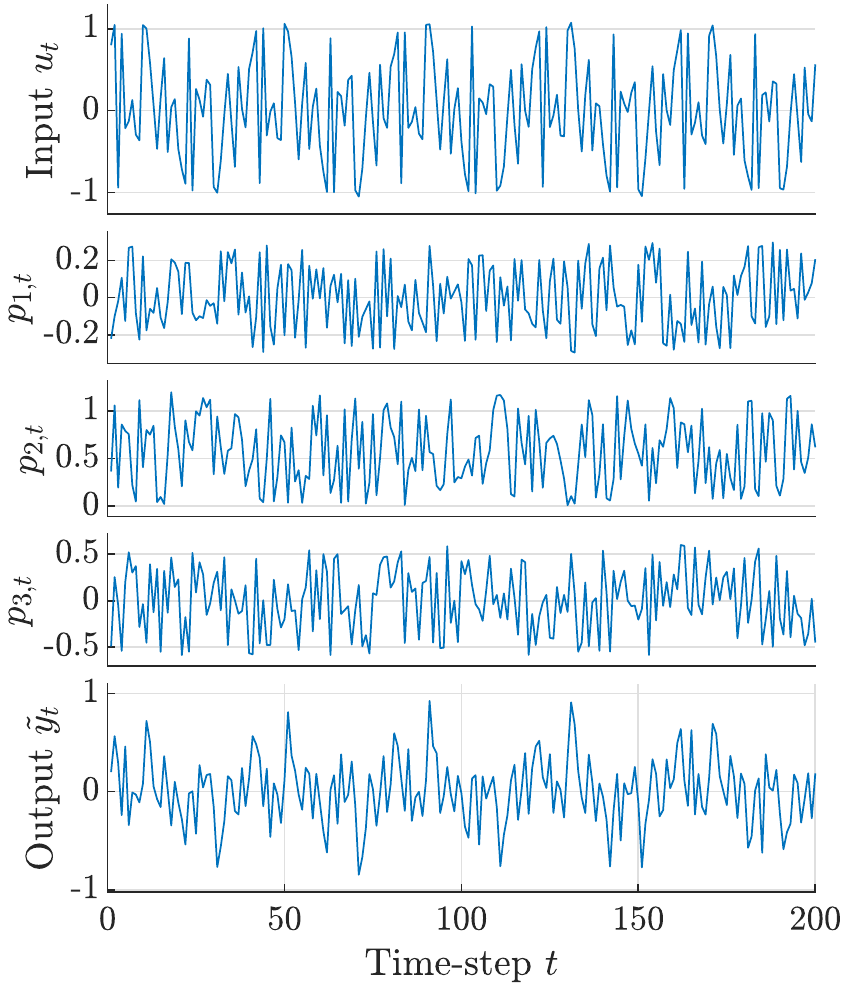}
       \caption{\testa data-set}
       \label{fig:data3}
    \end{minipage}\hfill
    \begin{minipage}[t]{0.245\linewidth}
       \includegraphics[width=\linewidth]%
       {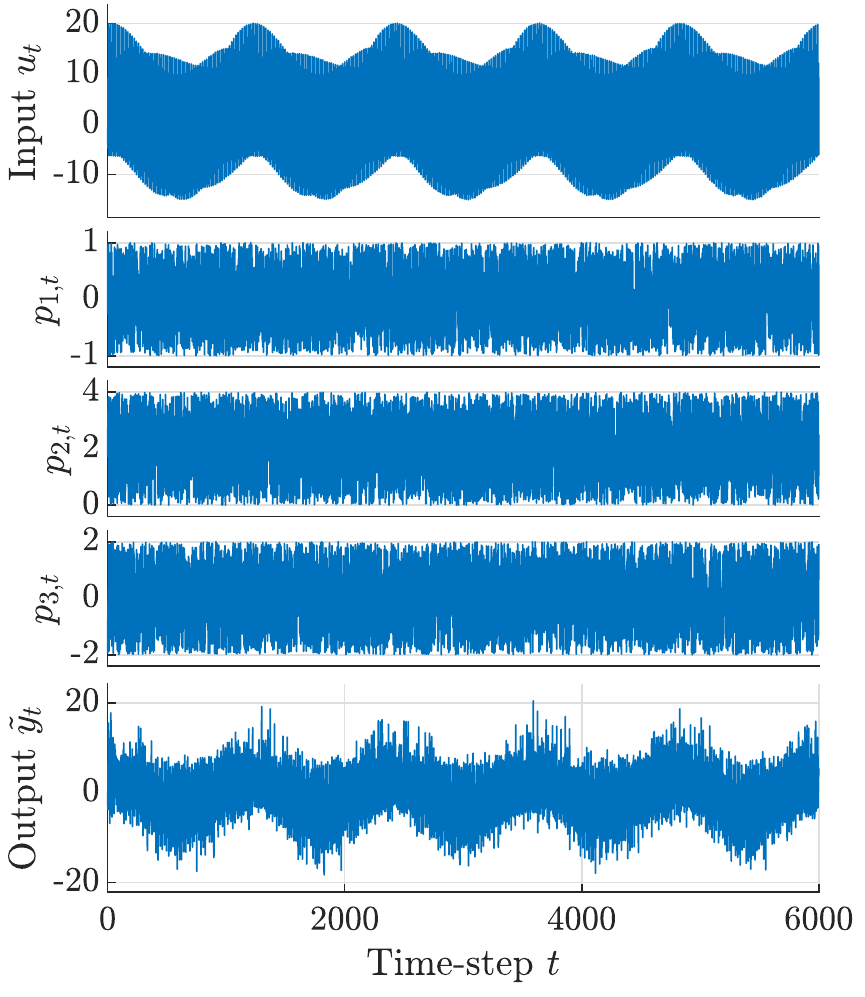}
       \caption{\testb data-set}
       \label{fig:data4}
    \end{minipage}%
\end{figure*}
With the following example\footnote{The data-sets and code used for this example can be found at \texttt{https://tinyurl.com/robstablpv}.}, we aim to demonstrate the effectiveness of the proposed robust and stable LPV-SS parametrization for deep-learning-based identification by comparing the training results with these models to the training results under a %
general LPV model structure. %
\subsection{Data-generation}
The data-generating system is considered to be in an LPV-SS form with output noise:
\begin{subequations}\label{eq:testsys}
\begin{align}
    x_{t+1} & = A^\mathrm{d}(p_t) x_t + B^\mathrm{d}(p_t) u_t, \\
    \tilde y_t & = C^\mathrm{d}(p_t)x_t + D^\mathrm{d}(p_t) u_t + e_t,
\end{align}
\end{subequations}
where, at time $t\in\mathbb{N}$, $u_t\in\R$ is the input, $p_t\in\R^3$ is the scheduling,  $x_t\in\R^3$ is the state, and $\tilde y_t\in\R$ is the output that is disturbed by an i.i.d. white noise signal $e_t\sim\mathcal{N}(0,0.08)$. The matrices $A^\mathrm{d},\dots,D^\mathrm{d}$ have static-affine dependence on $p_t$, i.e., $A^\mathrm{d}(p_t),\dots, D^\mathrm{d}(p_t)$ are of the form \( X(p_t) = X_0 + \sum_{i=1}^{\dnp}X_i p_{i,t} \) with %
\begingroup\allowdisplaybreaks
\begin{align*}
    A_0^\mathrm{d} & = \begin{bsmallmatrix} 
   -0.3885 & -0.1912 &  0.1631 \\
    0.3261 & -0.2583 & -0.9150 \\
   -0.1664 & -0.1384 &  0.0768
   \end{bsmallmatrix}, &&& B_0^\mathrm{d} &= \begin{bsmallmatrix}
   -3.4269 \\
   -0.3316 \\
   -2.1006 \end{bsmallmatrix}, \\   
   A_1^\mathrm{d} &= \begin{bsmallmatrix}
    0.2650 & -0.2214 & -0.1866 \\
    0.1747 &  0.1687 & -0.5876 \\
   -0.0477 & -0.1313 &  0.2863
   \end{bsmallmatrix}, &&&B_1^\mathrm{d} &= \begin{bsmallmatrix}
   -1.1096 \\
   -0.8456 \\
   -0.5727 \end{bsmallmatrix},\\
   A_2^\mathrm{d} & = \begin{bsmallmatrix}
    0.1476 & 0.1390 & 0.0901 \\
   -0.1242 & 0.1903 & 0.4027 \\
    0.0403 & 0.0845 & 0.0971
    \end{bsmallmatrix}, &&& B_2^\mathrm{d}& = \begin{bsmallmatrix}
   -0.5587\\
    0.1784\\
   -0.1969\end{bsmallmatrix}, \\
   A_3^\mathrm{d} &= \begin{bsmallmatrix}
    0.1613 & -0.0998 & -0.1652 \\
    0.0349 &  0.0645 & -0.1630 \\
    0.0098 & -0.0529 &  0.0591
    \end{bsmallmatrix}, &&& B_3^\mathrm{d} &= 0_{3\times1}, \\
   C_0^\mathrm{d} &= \begin{bsmallmatrix} -0.2097 & 0.0607 & 0.1421 \end{bsmallmatrix}, &&& \hspace{-25pt}C_1^\mathrm{d}=C_2^\mathrm{d}&=C_3^\mathrm{d}= 0_{1\times3}, \\
   D_0^\mathrm{d} & = 0.3,\quad  D_1^\mathrm{d} = 0.01, \quad D_2^\mathrm{d} = 0,\hspace{-5pt} &&& D_3^\mathrm{d} & = 0.04.
\end{align*}
\endgroup
For this system, $A^\mathrm{d}(p_t)$ satisfies that the spectral radius of $A^\mathrm{d}(p_t)$ is less than~1 for $p_t\in[-1, 1]\times[0, 4]\times[-2, 2]=\mathbb{P}$, which is considered as the scheduling range.

From \eqref{eq:testsys}, four data-sets are obtained: one \train and \valid data-set and two test-sets; \testa and \testb. We have generated these sets by applying an input to \eqref{eq:testsys} that is constructed with a white noise-signal with variance 0.05 added to a multi-sine. The multi-sine signal contains 10 sinusoidal components evenly distributed over the full normalized frequency spectrum. The scheduling signal is taken as a white noise with a uniform distribution over $\mathbb{P}$. The data-sets are composed of $N_\mathrm{b}$ trajectories, each of length $T$. The generated data-sets and their individual length-$T$ trajectories are uncorrelated. The specific details for the generated data-sets are listed in Table~\ref{tab:datasets}.
\begin{table}
    \centering
    \caption{Specifications of the generated data-sets}\label{tab:datasets}
    \begin{tabular}{c|cccc}
        Item~\textbackslash~Data-set & \train & \valid & \testa & \testb \\\hline
        Range $ u_t$ & $[-1, 1]$ & $[-1, 1]$ & $[-1, 1]$ & $[-20, 20]$ \\
        Range $ p_t$ & $0.3\mathbb{P}$ & $0.3\mathbb{P}$ & $0.3\mathbb{P}$ & $\mathbb{P}$ \\
        $T$ & $200$ & $200$ & $200$ & $6000$ \\ 
        $N_\mathrm{b}$ & $3200$ & $1280$ & $30$ & $1$ 
    \end{tabular}
    \vspace{-3mm}
\end{table}
Hence, data-set \testb is excited by and scheduled with an input and scheduling that are \emph{outside} the range represented in the \train and \valid data-sets. The generated data-sets are shown in Figs.~\ref{fig:data1}--\ref{fig:data4}. We want to highlight that with the aforementioned specification on the output-noise $e_t$, the \emph{signal-to-noise ratio} (SNR) for the \train, \valid and \testa data-sets is 12 dB. This implies that the lowest possible \emph{normalized root-mean-square error} (NRMSe) that we can achieve when simulating the trained models is approximately 25\%.
\subsection{Considered model structures}
 To learn, i.e., identify, \eqref{eq:testsys}, we consider the $\gamma$-Lipschitz LPV-SS model parametrization of Thm.~\ref{thm:lipschitz} with the following hyperparameters: The state-dimension of the $\gamma$-Lipschitz LPV-SS model is chosen as $\dnx=3$. The mapping $\phi_{\tilde{\theta}}:p\mapsto (X, Y,b)$ according to Thm.~\ref{thm:lipschitz} is chosen as a feedforward neural network for each component with 2 hidden layers, each with 50 ReLU activation neurons and a linear in- and output layer (note that $Z$ is empty). The value for $\gamma$ is set to~1, such that the model is ensured to have a Lipschitz bound of~1. Note that it is always possible to perform a hyperparameter optimization for $\gamma$ to improve the performance of the model.

The results of the identification with the $\gamma$-Lipschitz LPV-SS model are compared to estimation of an LPV model given by the following \emph{linear fractional representation} (LFR):
\begin{align}
    \left[\begin{array}{c}
	\!\!\!x_{t+1}\!\!\!\! \\\hdashline \!\!z_t\!\! \\ \!\!y_t\!\! \end{array}\right]\!\! &=\!\!\left[\begin{array}{c:cc}
	\!\!A(p_t)\! \!&\!\! B_{\mathrm{w}}(p_t)\!\! & \!\!B_{\mathrm{u}}(p_t)\!\!\\\hdashline
	\!\!C_{\mathrm{z}}(p_t)\! \!& \!\!0 \!\!&\!\! D_{\mathrm{zu}}(p_t) \!\!\\
	\!\!C_{\mathrm{y}}(p_t)\! \!&\! \!D_{\mathrm{yw}}(p_t) \!\!&\!\! D_{\mathrm{yu}}(p_t)\!\!
	\end{array}\right]\!\!\left[\begin{array}{l}
	\!\!x_t\!\!\! \\\hdashline \!\! w_t\!\!\! \\ \!\! u_t \!\!\! \end{array}\right] \!+\! \begin{bmatrix} b_x(p_t)\\b_z(p_t)\\b_y(p_t) \end{bmatrix},\notag\\
	w_t &=\sigma\left(z_t\right),\label{eq:LFRtotal}
\end{align}
where $\sigma:\R^{n_\mathrm{z}}\to\R^{n_\mathrm{w}}$ is a ReLU activation function and $A,\dots, D_\mathrm{yu}$ have affine dependence on $p_t$. Note that the data-generating system \eqref{eq:testsys} is contained in the model structure corresponding the LPV-LFR model \eqref{eq:LFRtotal}. For this model, we choose the following hyperparameters: The state-dimension is chosen as $\dnx=3$. The dependency of the matrices in~\eqref{eq:LFRtotal} on $p_t$ is, as aforementioned, static-affine. The dimension of $w_t$ and $z_t$ is 100, which implies that the corresponding NN component has one hidden layer with 100 neurons. %
The models are initialized randomly with matrices that have entries between $-0.1$ and $0.1$. %

\subsection{Training of the models}

We choose Adam \cite{kingma2014adam} as the optimizer with a learning-rate of $10^{-2}$ to minimize the loss function $\mathcal{L}$ in \eqref{eq:learning}. During training (similarly for validation), we forward simulate the models for the trajectories in the \train data-set, i.e., for 200 steps, initialized with a random initial condition taken from a normalized uniform distribution. To eliminate the effect of the wrongly chosen initial condition, we compute the simulation loss %
from $t=10$. The models are trained for 20 epochs with a batch size of $N_\mathrm{b}$ during optimization.

\subsection{Comparison of the results}
After training the models, we forward simulated them on both the \testa and \testb data-sets, again initialized with a random initial condition taken from a normalized uniform distribution. To assess the correctness of the training result, we measure the simulation accuracy using the NRMSe, i.e.,
\[ \mathrm{NRMSe}(\tilde{y},y) = \frac{1}{\dny}\sum^{\dny}_{i=1}\frac{\big({\tfrac{1}{T}\sum_{t=1}^T(\tilde{y}_{i,t}-y_{i,t})^2}\big)^{\tfrac{1}{2}}}{\mathrm{std}(\tilde{y}_i)},\]
with $\mathrm{std}(\tilde{y})$ the sample standard deviation of the measured output sequence $\tilde{y}$. The simulation results for both models on data-set \testa are shown in Fig.~\ref{fig:ressec},
while the simulation responses for both models on data-set \testb are shown in Fig.~\ref{fig:resfull}.
In Fig.~\ref{fig:ressec}, we only show the simulation result that had the lowest NRMSe out of the 30 responses coming from \testa. The average NRMSe over the 30 trajectories is 0.4507 for the $\gamma$-Lipschitz LPV-SS model and 0.3378 for the LPV-LFR model. 

The results in Fig.~\ref{fig:ressec} show that after 20 epochs of training, the LPV-LFR model resulted in a better prediction model for the data-generating system \eqref{eq:testsys} in terms of the NRMSe. This can be caused by a too conservative choice of $\gamma$ for the $\gamma$-Lipschitz model to represent the actual dynamics and the more simple parametrization of the LPV-LFR compared to the $\gamma$-Lipschitz model, which could result in a faster convergence during optimization. %

The strength of the $\gamma$-Lipschitz model comes forward when we simulate the trained models on the \testb data-set. As can be observed in Fig.~\ref{fig:resfull}, the response of the $\gamma$-Lipschitz model to a scheduling that goes beyond the range that was seen during training still respects the Lipschitz bound of~1, while the output response of the LPV-LFR model explodes in terms of magnitude. In fact, although not further discussed in this paper, when we identify an LPV-SS model using state-of-the-art LPV identification methods \cite{CoxToth2021, lpvcore}, we obtain unstable behavior when simulated on \testb. As highlighted in Section~\ref{s:introduction}, this convenient property of the $\gamma$-Lipschitz LPV-SS model makes this parametrization attractive to use for modeling problems in, e.g., the process industry, where the data-generating system is often dependent on many measurable exogenous parameters. In these situations, experiment-design is often limited in terms of excitation due to cost, while, during normal operation, the true underlying system (e.g., reactor) is fed with inputs far outside the excitation range of the experiment. The guaranteed Lipschitz property ensures that the trained LPV-SS model will not behave unexpectedly when simulated with the typical inputs.
\begin{figure}
    \centering
    \begin{subfigure}[t]{\linewidth}
       \includegraphics[scale=0.9, trim=0mm 0.95mm 0mm 0.7mm, clip]{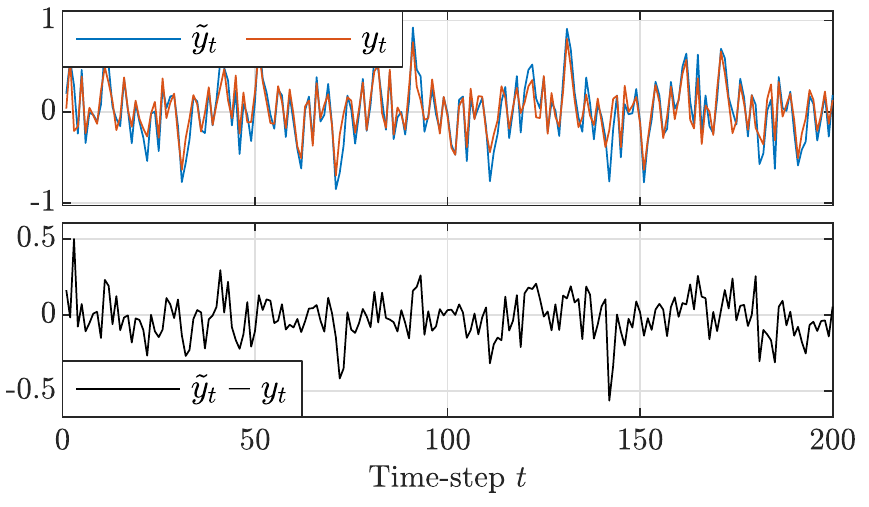} \vspace{-1mm}
       \caption{$\gamma$-Lipschitz LPV-SS model (NRMSe: 0.4179)} \vspace{1mm}%
       \label{fig:ressec1}
    \end{subfigure}
     \begin{subfigure}[t]{\linewidth}
       \includegraphics[scale=0.9, trim=0mm 0.95mm 0mm -0.3mm, clip]{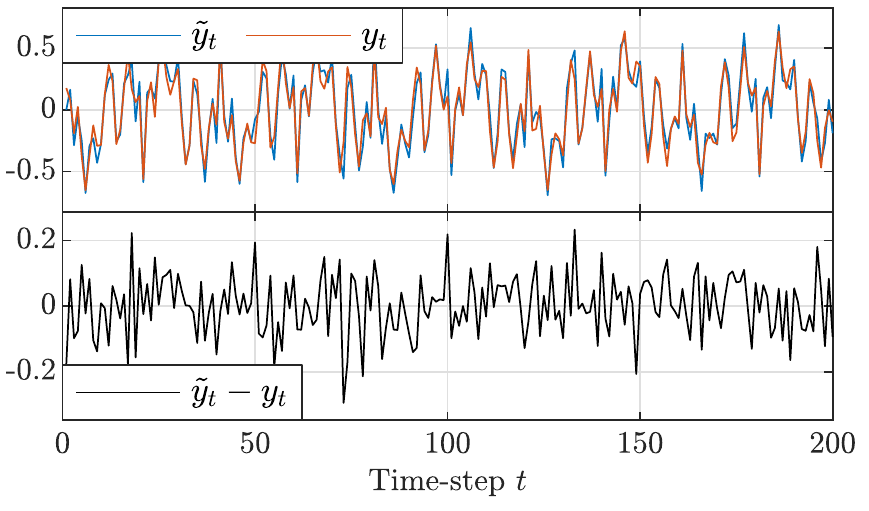} \vspace{-1mm}
       \caption{LPV-LFR model (NRMSe: 0.2910)}%
       \label{fig:ressec2}
    \end{subfigure}
    \caption{Simulation results on the \testa data-set, with $\tilde{y}$ the output in the data-set, and $y$ the predicted output of the trained models.}\label{fig:ressec}
\end{figure}
\begin{figure}
    \centering
    \begin{subfigure}[t]{\linewidth}
       \includegraphics[scale=0.9, trim=0mm 0.95mm 0mm 0.7mm, clip]{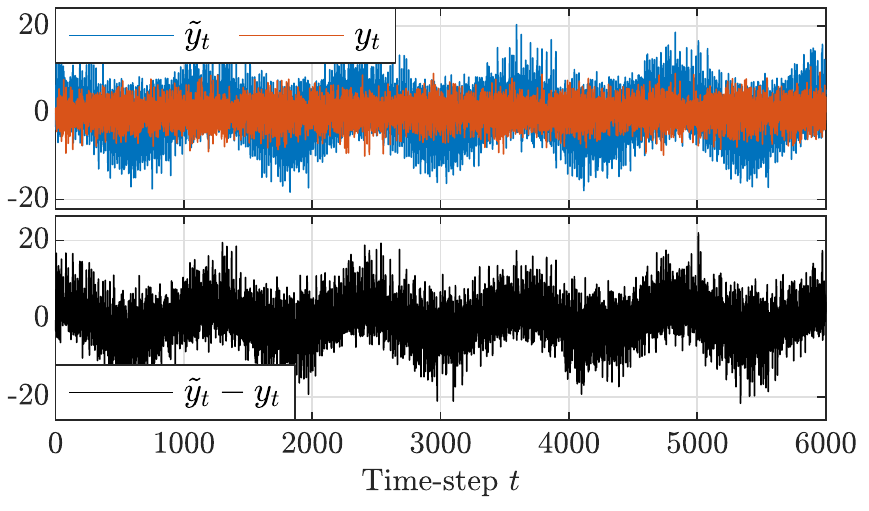} \vspace{-1mm}
       \caption{$\gamma$-Lipschitz LPV-SS model (NRMSe: 1.0227)} \vspace{1mm}
       \label{fig:resful1}
    \end{subfigure}
     \begin{subfigure}[t]{\linewidth}
       \includegraphics[scale=0.9, trim=0mm 0.95mm 0mm 0.0mm, clip]{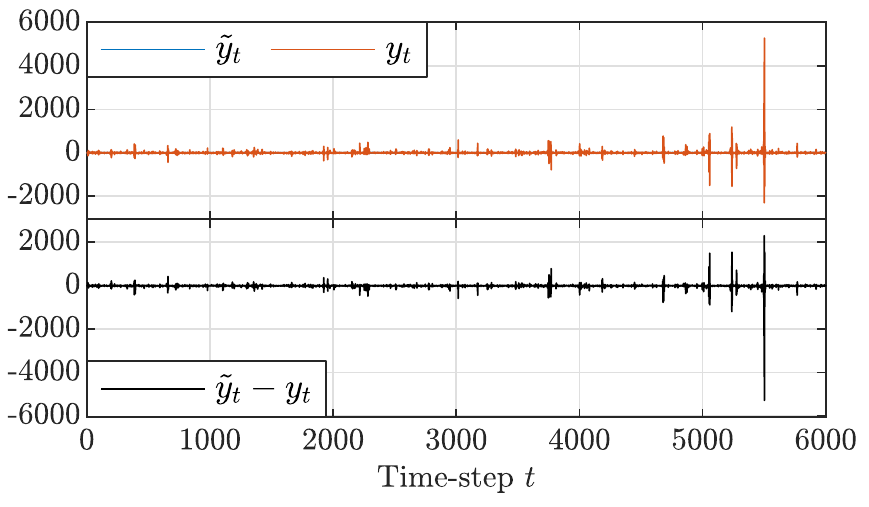} \vspace{-1mm}
       \caption{LPV-LFR model (NRMSe: 20.599)}
       \label{fig:resful2}
    \end{subfigure}
    \caption{Simulation results on the \testb data-set, with $\tilde{y}$ the output in the data-set, and $y$ the predicted output of the trained models.}\label{fig:resfull}
\end{figure}

\section{Conclusions}\label{s:conclusions}
This work introduces stable and robust parametrizations of LPV state-space models based on the Cayley transform. By means of contraction theory, we can \emph{a priori} guarantee \emph{global} stability and performance (in terms of $\gamma$-Lipschitz) properties of the to-be-trained model. The proposed model parametrizations are highly flexible and require no further constraints or optimization based stability checks compared to alternative solutions. %
The strength of having these guaranteed properties is demonstrated in an example that considers an LPV system-identification problem.

\appendix
The proofs of Thms.~\ref{thm:contracting}~and~\ref{thm:lipschitz} make use of the following lemmas. \arXver{See \cite{arxivversion} for the proofs of these lemmas.}{}
\begin{lemma}\label{lem:cayley}
Let $M \in \R^{n\times m}$ with $n\geq m$. Then, $M^\top M\prec I$ if and only if there exist $X,Y\in \R^{m\times m}$ and $Z\in \R^{(n-m)\times m}$ such that 
\begin{equation}
    M=\begin{bmatrix}
        \cayley(N) \\
        -2 Z(I+N)^{-1}
    \end{bmatrix}
\end{equation}
where $N=X^\top X+Y -Y^\top +Z^\top Z+\epsilon I$.
\end{lemma}
\arXver{}{
\begin{proof}
{\bf Sufficiency.} Both $I+N$ and $I+N^\top$ are invertible as $N^\top +N=2(\epsilon I +X^\top X+Z^\top Z)\succ 0$. Therefore, $M$ is well-defined and satisfies
\begin{multline}\label{eq:cayley-ext}
        (I+N^\top)(I+N)-(I+N^\top)M^\top M(I+N) \\
        = (I+N^\top)(I+N) - (I-N^\top)(I-N)-4Z^\top Z\\
        = 2(N^\top +N)-4Z^\top Z=4(\epsilon I + X^\top X)\succ 0,
\end{multline}
which implies that $M^\top M\prec I$.

{\bf Necessity.} First, we partition $M$ by $M^\top=\begin{bmatrix}
    M_1^\top & M_2^\top
\end{bmatrix}$. Then, $I+M_1$ is invertible since $M_1^\top M_1+M_2^\top M_2 \prec I$. From \eqref{eq:cayley-ext}, we have 
\begin{equation}
    N=\cayley(M_1),\quad Z=-\tfrac{1}{2}M_2(I+N).
\end{equation}
Let $H:=\tfrac{1}{2}(N^\top+N)-Z^\top Z$. We can further obtain that
\begin{align*}
    H&=\begin{multlined}[t]
            \tfrac{1}{2}(I+M_1)^{-\top}(I-M_1^\top)(I+M_1)(I+M_1)^{-1} + \\
        \tfrac{1}{2}(I+M_1^\top)^{-1}(I+M_1^\top)(I-M_1)(I+M_1)^{-1} - \\
         (I+M_1)^{-\top}M_2^\top M_2(I+M_1)^{-1}
        \end{multlined} \\
    &= (I+M_1)^{-\top}(I-M_1^\top M_1 -M_2^\top M_2)(I+M_1)^{-1}\succ 0.
\end{align*}
Then, we can choose a sufficiently small $\epsilon>0$ such that $\hat{H}=H-\epsilon I\succeq 0$. By taking the SVD decomposition $\hat{H}=U^\top \Sigma U$, we can construct $X,Y$ as follows
\begin{equation}
    X=\Sigma^{\frac{1}{2}}U,\quad Y=\tfrac{1}{2}N.
\end{equation}
Substituting $X,Y,Z$ into \eqref{eq:cayley-ext} recovers the matrix $M$.
\end{proof}
}

\begin{lemma}\label{lem:cayley-2}
    Let $M$ be a square matrix that does not have an eigenvalue of $-1$. Then, $M^\top M=I$ if and only if there exists a square matrix $Y$ such that $M=\cayley(Y-Y^\top)$.
\end{lemma}
\arXver{}{
\begin{proof}
{\bf Sufficiency.} By defining $N:=Y-Y^\top$ we have 
\begin{align*}
    M^\top M &= (I+N^\top)^{-1}(I-N^\top)(I-N)(I+N)^{-1} \\
    &=(I+N^\top)^{-1}(I-N^\top-N+N^\top N)(I+N)^{-1} \\
    &=(I+N^\top)^{-1}(I+N^\top+N+N^\top N)(I+N)^{-1} \\
    &=(I+N^\top)^{-1}(I+N^\top)(I+N)(I+N)^{-1} = I.
\end{align*}
{\bf Necessity.} Since $-1$ is not an eigenvalue of $M$, we have that $I+M$ is invertible and thus $N=\cayley(M)$ is well-defined. Then, we can verify that $N$ is asymmetric as
\begin{align*}
    N^\top + N &= (I+M^\top)^{-1}(I-M^\top)+(I-M)(I+M)^{-1}\\
&=2(I+M^\top)^{-1}(I-M^\top M)(I+M)^{-1}=0.
\end{align*}
By taking $Y=\tril(N)$, %
we have $N=Y-Y^\top$.
\end{proof}
}

\bibliographystyle{IEEEtran}
\bibliography{refs}

 \end{document}